\documentclass[a4paper,11pt]{article}
\usepackage{pos}
\usepackage{multicol}
\usepackage{subfigure}
\usepackage{hyperref}
\hypersetup{colorlinks=true,urlcolor=violet,anchorcolor=blue,citecolor=blue,filecolor=blue,linkcolor=violet,menucolor=blue, linktocpage=true,pdfproducer=medialab}
\usepackage{graphicx}
\usepackage{float}

\newcommand{\be}{\begin{equation}}
\newcommand{\ee}{\end{equation}}
\newcommand{\ba} {\begin{equation}\begin{aligned}}
\newcommand{\ea} {\end{aligned}\end{equation}}

\newcommand{\sL}{\mathscr{L}}

\newcommand{\hc}{\text{h.c.}}

\newcommand{\ov}[1]{\overline{#1}}
\newcommand{\sign}[1]{\text{sgn}(#1)}

\def\vs{{\textit vs.} }

\def\tH{\widetilde{H}}

\newcommand{\TeV}{\ \text{TeV}}
\newcommand{\GeV}{\ \text{GeV}}

\newcommand{\PQ}{U(1)_\text{PQ}}
\newcommand{\LN}{U(1)_\text{LN}}

\DeclareUnicodeCharacter{2212}{-}

\graphicspath{{Plots/}}

\title{GeV ALP from TeV Vector-like leptons}

\author[a]{Arturo de Giorgi}
\author*[b]{Marta Fuentes Zamoro}
\author[b]{Luca Merlo}

\affiliation[a]{Institute for Particle Physics Phenomenology, Department of Physics, Durham University,\\
Durham, DH1 3LE, U.K.}
  \affiliation[b]{Departamento de Física Teórica and Instituto de Física Teórica UAM/CSIC, \\
  Universidad Autónoma de Madrid,  Cantoblanco, 28049, Madrid, Spain}

\emailAdd{arturo.de-giorgi@durham.ac.uk}
\emailAdd{marta.zamoro@uam.es}
\emailAdd{luca.merlo@uam.es}

\abstract{We present a model where a GeV axion-like-particle (ALP) 
is predicted in a large portion of the parameter space due to the presence of explicit Peccei-Quinn symmetry-breaking terms in an exotic leptonic sector. The latter provides a solution to the muon $g-2$ anomaly, within the framework of the Linear Seesaw neutrino mechanism. The spectrum is extended by a complex scalar singlet only transforming under the Peccei-Quinn symmetry, which generates the ALP. Its couplings with fermions can continuously span over many orders of magnitude, which constitutes a specific feature of this model in contrast to generic ultraviolet constructions. Interestingly, these couplings are suppressed by the ALP characteristic scale that can be as low as the TeV scale, which represents a novel feature of the model and opens up to several phenomenological consequences.}

\FullConference{2nd Training School and General Meeting of the COST Action COSMIC WISPers  (CA21106) (COSMICWISPers2024)\\
 10-14 June 2024 and 3-6 September 2024\\
Ljubljana (Slovenia) and Istanbul (Turkey)\\}
\begin{document}
\maketitle
\section{Introduction}
The Standard Model (SM), despite being one of the most predictive theories present, has some open puzzles that require further understanding on our side, such as neutrino masses or the strong CP problem. The axion is one of the most promising candidates to explain the latter one. Related to it, an Axion-like Particle (ALP) is a pseudo-scalar with dominant derivative couplings, whose mass is not strictly related to the scale $f_a$ thus opening up the parameter space that can be studied. Expanding on the work done in \cite{Luca}, we here summarize the work done in \cite{degiorgi}, where an ALP, coming from the SSB of $\PQ$, plays a fundamental role in the phenomenology presented here. 
\section{Model description}
\begin{multicols}{2}
We work with a specific particle spectrum that includes all SM gauge bosons and fermions, while including modifications only to the muon and muonic neutrinos' interactions. To this end, we introduce two heavy RH neutrinos, $N_R$ and $S_R$, as well as a vector-like electroweak lepton doublet (so far, this particle content is the one in \cite{Luca}). On top of this, we expand it by incorporating a complex scalar field $\phi$, which is a singlet under the SM and is associated to a $\PQ$, which will be spontaneously broken. For an overview of the particle content and its representation under the SM gauge groups as well as the $\PQ$, see Table \ref{table:particlespectrum}.
\begin{table}[H]
	\centering
	\begin{tabular}{c||c|c||c|c|}
		&$SU(2)_{L}$&$U(1)_{Y}$ &$U(1)_{L}$ & $\PQ$\\
		\hline \hline
		$\ell_L$&$\bf{2}$&$-1/2$&$1$&$n_{N_R}$\\
		$\mu_{R}$&$1$&$-1$&$1$&$n_{\psi_L}$\\
		$H$&$\bf{2}$&$1/2$&$0$&$0$\\
		\hline
		$N_{R}$&$1$&$0$&$1$&$n_{N_R}$\\
		$S_{R}$&$1$&$0$&$-1$&$n_{S_R}$\\		
		\hline	
    		$\psi_{L}$&$\bf{2}$&$-1/2$&$1$&$n_{\psi_L}$\\
		$\psi_{R}$&$\bf{2}$&$-1/2$&$1$&$n_{\psi_R}$\\
		\hline
		$\phi$&$1$&$0$&$0$&$n_\phi$\\
		\hline		
	\end{tabular}
	\caption{\em Transformation properties of the SM leptons $\ell_{L}$ and $\mu_{R}$, the Higgs doublet H, the HNLs $N_R$ and $S_R$, the vector-like EW lepton doublet $\psi$ and the scalar $\phi$ under $SU(2)_L\times U(1)_{Y}\times\LN\times \PQ$.}
	\label{table:particlespectrum}
\end{table}
\end{multicols}
The part of the Lagrangian relevant for the phenomenology studied here is the part responsible for giving masses to the second SM lepton generation and the exotic leptons. We will assume the following expression for it
\begin{equation}
	\begin{aligned}
		 -\sL_Y=&\phantom{+}Y_N\ov{\ell_L} \tH N_R+Y_R \ov{\psi_L} H \mu_R+\\
		&+\delta_{x,0} \Lambda \overline{N_R^c} S_R+\delta_{|x|,1} \alpha_N \phi^{(\ast)} \overline{N_R^c} S_R+\delta_{y,0} M_\psi \ov{\psi_L} \psi_R+\delta_{|y|,1} \alpha_\psi\phi^{(\ast)} \ov{\psi_L} \psi_R+ \\
		& +Y_V \overline{S_R^c} \tH^{\dagger} \psi_R+Y_{V^\prime} \ov{\psi_L} \tH N_R +\epsilon Y_S\ov{\ell_L} \tH  S_R+\hc\,, \\
	\end{aligned}
	\label{lmodel}
\end{equation}
where $\tH\equiv i \sigma_2 H$, being $\sigma_2$ the second Pauli matrix, and $\delta_{i,j}$ is the Kronecker delta. The quantities $Y_N$, $Y_R$, $Y_V$, $Y_{V'}$, $Y_S$, $\alpha_N$ and $\alpha_\psi$ are dimensionless parameters and $\Lambda$ and $M_\psi$ are instead masses. On the other hand, $x$ and $y$ are variables that can only acquire three values, either $0$ or $\pm1$, distinguishing in this way four different realisations, apart from the one already discussed in \cite{Luca} (for a full analysis regarding them, please refer to \cite{degiorgi}). Finally, $\epsilon$ is a small parameter and is responsible for the active neutrino mass {\it \`a la} LSS mechanism. 
In the rest of this proceeding we will focus on "Model B", which is characterised by the dynamical generation of the mass related to $\psi$, while the mass $\Lambda$, related to $N_R$ and $S_R$ is a free parameter of the Lagrangian. Unless otherwise stated, $\alpha_\psi=1$, so that $M_\psi=f_a/\sqrt{2}$.
\subsection{Lepton mass}
After the Higgs and the scalar singlet acquire a VEV, the mass Lagrangian for the neutral and charged sectors reads 
\be
\begin{aligned}
    -\sL_Y\supset \frac{1}{2}\ov{\chi}\mathcal{M}_\chi \chi^c+ \ov{\zeta_L}\mathcal{M}_\zeta \zeta_R+\hc\,,
\end{aligned}
\ee
where we defined the neutral lepton multiplet $\chi$ and the charged one $\zeta$ as
\begin{equation}
    \chi\equiv (\nu_L, N_R^c,S_R^c,\psi_L^0,\psi_R^{0\,c})^T,\qquad \zeta\equiv (\mu,\psi^-)^T\,,
    \label{multiplet}
\end{equation}
and the charge conjugation operation is defined by $N^c_R \equiv \mathcal{C} \ov{N_R}^T$ with $\mathcal{C} $ being the charge conjugation matrix. $M_\chi$ and $M_\zeta$ are the neutral and charged mass matrices, respectively.
It is relevant to note that there is no direct muon mass term in the Lagrangian of Eq.\ref{lmodel}. Therefore, the muon mass must be generated at loop-level. After considering wave function renormalization effects, the muon pole-mass at 1-loop reads
\be
\hat{m}_\mu=\delta m_\mu=-\dfrac{m_N\,m_R\,\Lambda}{8\,\pi^2\,v^2}
    \left(\dfrac{m_V}{M_\psi}+\dfrac{m_{V'}}{\Lambda}\right)
    \left[1+\dfrac{1}{M_\psi^2-\Lambda^2}\left(M^2_\psi
    \log{\dfrac{\mu_\text{R}^2}{M_\psi^2}}-
    \Lambda^2
    \log{\dfrac{\mu_\text{R}^2}{\Lambda^2}}\right)\right]\,,
\label{deltammu}
\ee
where we neglected $\mathcal{O}(\Lambda^{-2},M_{\psi}^{-2})$ corrections.
\subsection{ALP Lagrangian}
\paragraph{ALP mass}
The existence of explicit $\PQ$ breaking terms in Eq. \ref{lmodel} generates a radiative mass for the ALP after SSB, which can be obtained using the Coleman-Weinberg potential in the $\ov{\text{MS}}$-scheme considering only the neutral sector (it involves all explicit PQ breaking). The expression reads
{\footnotesize
\begin{equation}
        f_a^2 m_a^2=\dfrac{(\ov{\delta}_{x,1}+\ov{\delta}_{y,1})^2}{4\pi^2}\left(\dfrac{m_V m_{V'}\Lambda M_\psi}{M_\psi^2-\Lambda^2}\right)\left[\dfrac{(M_\psi^2+\Lambda^2)}{2}\log\left(\dfrac{M_\psi^2}{\Lambda^2}\right)+(M_\psi^2-\Lambda^2)\left(\log\left(\dfrac{M_\psi\Lambda}{\mu_R^2}\right)-1\right)\right]\,,
    \label{eq:massALPatLO}
\end{equation}}
neglecting terms $\mathcal{O}(\Lambda^0,M_\psi^0)$, where we defined a Kronecker-delta with sign $\ov{\delta}_{x,1}\equiv \sign{x}\delta_{|x|,1}$, $\ov{\delta}_{y,1}\equiv\sign{y}\delta_{|y|,1}$. An interesting point to note in the above expression is that it vanishes for $M_\psi=\Lambda$, thus the NLO contribution would be necessary.\\

It is also relevant to study how the ALP interacts with the different particles present in the SM.
\paragraph{ALP-Gauge couplings}
In order for the ALP to interact with the gauge bosons, a non-vanishing chiral charge for $\psi$ is required, which means that $M_\psi$ must be generated dynamically. After doing the full 1-loop calculation, we see that 
\begin{align}
    &g_{a\gamma\gamma}=\ov{\delta}_{y,1}\frac{\alpha_\text{em}}{\pi f_a}\,,&&g_{aZZ}=\ov{\delta}_{y,1}\frac{\alpha_\text{em}}{6\pi f_a s_{2\theta_W}^2}\left(c_{4\theta_W}+7\right)\,,&&g_{aWW}=\ov{\delta}_{y,1}\frac{\alpha_\text{em}}{2\pi f_a s_{\theta_W}^2}\,.
    \label{eq:vector-coupling}
\end{align}
\paragraph{ALP-muon interaction} The ALP-muon coupling arises at 1-loop as the ALP does not couple to muons and no coupling at tree-level is allowed. After calculating the corresponding triangle diagrams we reach a coupling for the ALP with muons given by 
\begin{equation}
    g_{a\mu\mu}
    =\frac{(\ov{\delta}_{x,1}+\ov{\delta}_{y,1})}{f_a}\times\left(\frac{Y_V}{Y_V+\left(\frac{M_\psi}{\Lambda}\right) Y_{V'}}\right)\,,
    \label{eq:muon-coupling}
\end{equation}
which can span over several orders of magnitude.
\section{Phenomenological analysis}
\begin{figure}[h]
\centering
\subfigure[{ALP mass}\label{fig:massALP}]{\includegraphics[width=0.385\linewidth]{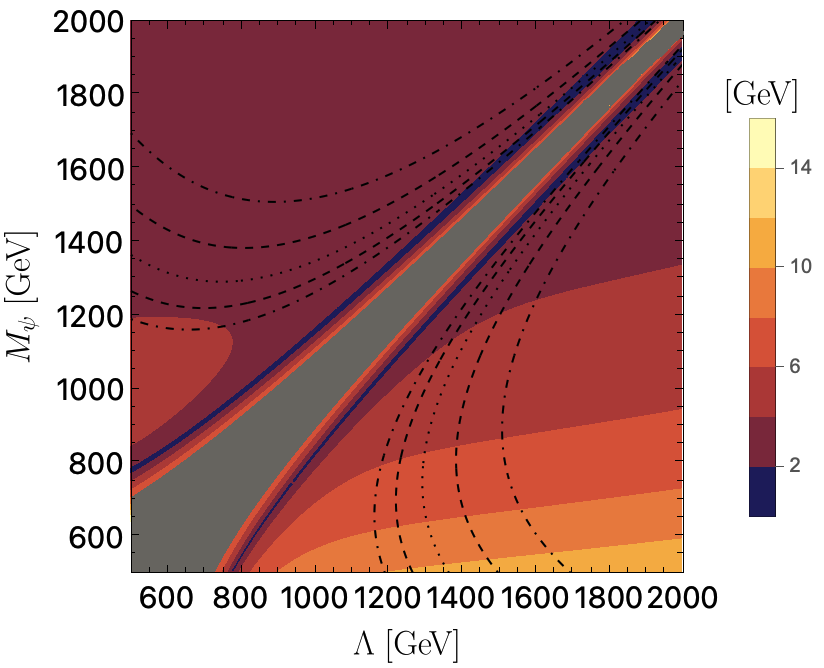}}
\subfigure[{Coupling to muons}\label{fig:muon-coupling}]{\includegraphics[width=0.385\linewidth]{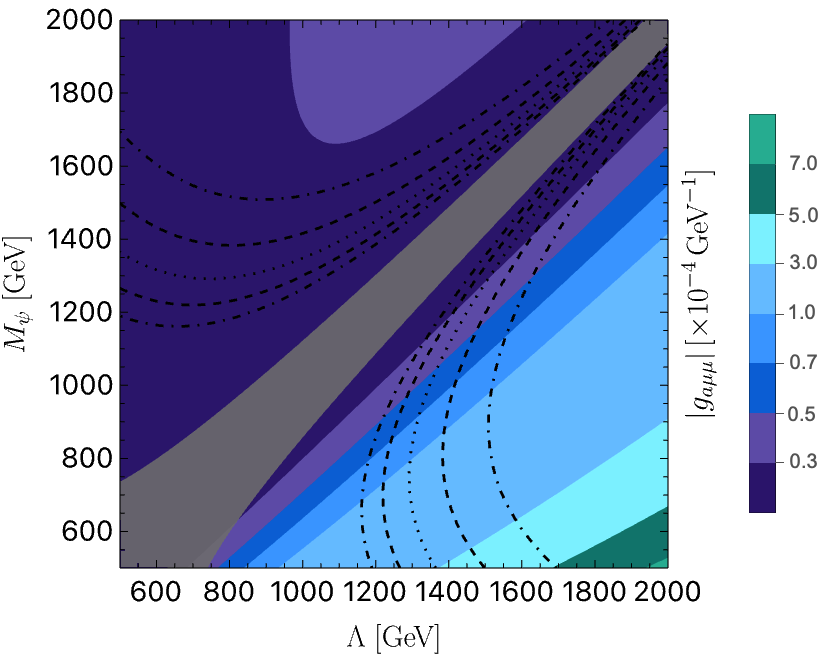}}
\caption{\em ALP mass and ALP coupling to muons the parameter space $\Lambda$ \vs $M_\psi$. In both plots $Y_V=0.1$. The grey-shaded region correspond to areas in which $ |Y_{V^\prime} |>5$ and thus perturbativity is not respected. The dotted lines represent the points in the parameter space in which the $(g-2)_\mu$ is resolved exactly, while the dashed(-dotted) lines include the region in which the $(g-2)_\mu$ is accounted for at $1\sigma(2\sigma)$. \textbf{Left:} ALP mass. \textbf{Right:} Values of the coupling of the ALP to muons in the studied parameter space.}
\label{fig:paramatlas}
\end{figure}

After studying all the theoretical implications of our model, it is enriching to see the parameter space we get by applying the constraints stemming from several experimental measurements, such as $m_W$ or the coupling of the $W,\, Z$ to muons; as well as the reproduction of the light neutrino masses. Choosing to reproduce the atmospheric mass splitting places a bound on $ |\epsilon Y_S Y_N/\Lambda |\sim 8.3\cdot 10^{-13} \TeV^{-1}$.
Additionally, one of the main objectives of the model proposed here is to shine light on the $(g-2)_\mu$ anomaly\footnote{To that purpose, we take the value $\delta a_\mu^\text{exp}\equiv a_\mu^\text{exp}-a_\mu^\text{SM}= (2.49\pm 0.49)\cdot 10^{-9}$. Recent lattice calculations contradict this value. Nevertheless, and for the time being, we will neglect such contributions. If in the future the results from lattice are confirmed, the $(g-2)_\mu$ would become one of the most stringent limits of this model.}. To that end, we calculated the ALP contribution to this observable and found that it was negligible. Therefore, the $(g-2)_\mu$ will be dominated by the EW contributions \cite{Luca}. It is possible to relate $\delta a_\mu$ with $m_\mu$, which allows us to know the value of $Y_{V^\prime}$ once $Y_V$ is fixed in the parameter space of $(M_\psi, \Lambda)$. In Fig.\ref{fig:paramatlas} we see the behaviour of the ALP mass and the coupling to muons in the case $Y_V$ = 0.1, where we have set $m_\mu=m_\mu^\text{exp}$.

\begin{figure}[h]
\centering
\includegraphics[width=0.8\linewidth]{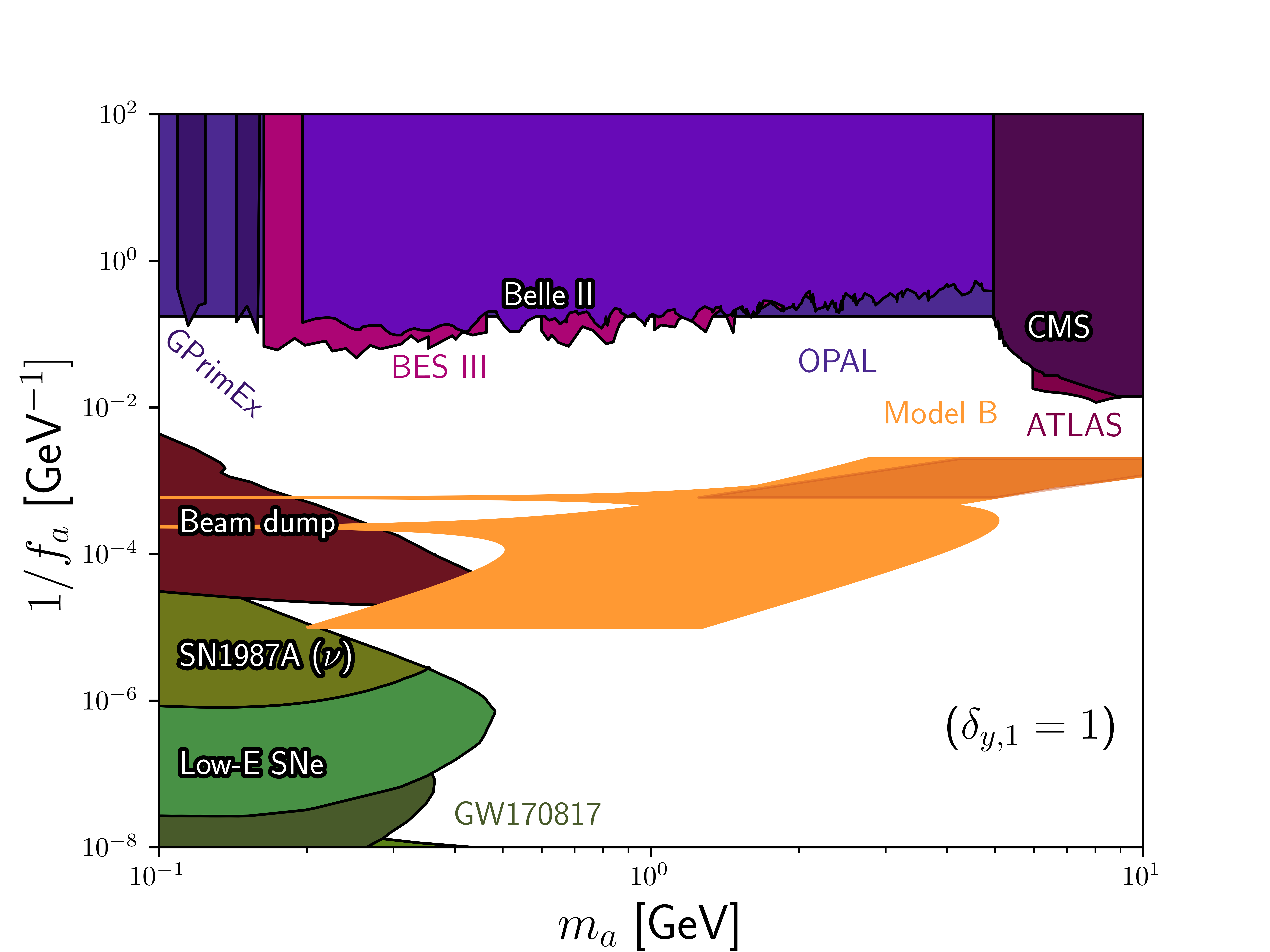}
\caption[]{\em Photon-ALP coupling as a function of $m_a$. Adapted from Ref.\cite{OHare}. The light orange region corresponds to the parameter space when $Y_{V,V^\prime}\subset\left[0.05,0.4\right]$ and $\alpha_{N,\psi}\subset\left[0.5,1.25\right]$. $\Lambda=1500 \GeV$. The darker orange region represents, instead, the benchmark point defined by $\Lambda=1500$, $M_\psi=600\GeV$ and the Yukawas $Y_{V,V^\prime}\subset\left[0.05,0.4\right]$, while $\alpha_{N,\psi}>0.5$.}
\label{fig:photonALP}
\end{figure}

It is also possible to extract bounds on $f_a$. From non-resonant ALP searches \cite{Bonilla}, we obtain $ |f_{aWW} |\geq 1.7\GeV$ and $ |f_{aZZ} |\geq 1.3 \GeV$. Additionally, we can extract bounds on $f_a$ from $g_{a\gamma\gamma}$, allowing us to compare $m_a$ vs $f_a$. In our case we show in Fig.\ref{fig:photonALP} the parameter space we would live in for the model B. As it can be seen, our model lives in a region of the parameter space which still has not been excluded by experimental searches, which means that this model can be tested in the upcoming future.

\section{Conclusions}
In this work, we present an UV completion with an exotic lepton sector that provides a realistic mass for active neutrinos via the linear low-scale seesaw mechanism. Additionally, it represents a viable solution to $(g-2)_\mu$. The expected mass range is within experimental reach, as the HNLs live in the TeV-scale and we obtain a mass of the ALP of $\mathcal{O}(\text{GeV})$ for a scale $f_a\sim \TeV$; thus making this model testable at colliders.
%
\section*{Acknowledgements}
\noindent The work of MFZ is supported by the grant PID2022-137127NB-I00 funded by MCIN/AEI/ 10.13039/501100011033 and by the Spanish MIU through the National Program FPU (grant number FPU22/03625).

\end{document}